\documentclass[authoryear,12pt]{elsarticle}
\pdfoutput=1

\usepackage[colorlinks=true,linkcolor=black, citecolor=blue, urlcolor=blue]{hyperref}
\usepackage{multirow}
\usepackage{rotating}
\usepackage{tabularx}
\usepackage{graphicx}
\usepackage[normalem]{ulem}
\useunder{\uline}{\ul}{}
\usepackage[utf8]{inputenc}
\usepackage{booktabs, makecell,dcolumn}
\usepackage[font=small, skip=8pt]{caption}
\usepackage[flushleft]{threeparttable}
\usepackage{adjustbox}
\usepackage{amsmath}
\usepackage{amsfonts}
\usepackage[capposition=top]{floatrow}
\usepackage{setspace}
\usepackage[top=1in, bottom=1in, left=1in, right=1in]{geometry}
\usepackage{footmisc}
\usepackage{todonotes}
\usepackage[]{algorithm2e}
\usepackage{verbatim} 
\usepackage[normalem]{ulem}  
\usepackage{yhmath} 
\usepackage[nice]{nicefrac} 

\usepackage{dirtytalk}  

\usepackage{multicol}  

\newcommand{\RR}{\mathbb{R}}

\newcommand{\NN}{\mathbb{N}}

\newcommand{\EE}{\mathbb{E}}

\newcommand{\fN}{\mathfrak{N}}
\newcommand{\fS}{\mathfrak{S}}

\newcommand{\tpi}{\texttt{i}}
\newcommand{\tps}{\texttt{s}}
\newcommand{\tpt}{\texttt{t}}
\newcommand{\tpw}{\texttt{w}}
\newcommand{\tpT}{\texttt{T}}

\newcommand{\Dt}{\Delta t}
\newtheorem{remark}{Remark}

\DeclareMathOperator*{\argmax}{arg\,max}

\DeclareMathOperator{\LR}{LR}
\DeclareMathOperator{\Var}{Var}

\makeatletter
    \def\ps@pprintTitle{%
      \let\@oddhead\@empty
      \let\@evenhead\@empty
      \let\@oddfoot\@empty
      \let\@evenfoot\@oddfoot
    }
\makeatother

\makeatletter
\renewcommand{\paragraph}{\@startsection{paragraph}{4}{0ex}%
    {-3.25ex plus -1ex minus -0.2ex}%
    {1.5ex plus 0.2ex}%
    {\normalfont\normalsize\itshape}}
\makeatother
 
\stepcounter{secnumdepth}
\stepcounter{tocdepth}

\begin{document}

\title{Randomized Signature Methods in Optimal Portfolio Selection}

\begin{abstract}
We present convincing empirical results on the application of Randomized Signature Methods for non-linear, non-parametric drift estimation for a multi-variate financial market. Even though drift estimation is notoriously ill defined due to small signal to noise ratio, one can still try to learn optimal non-linear maps from data to future returns for the purposes of portfolio optimization. Randomized Signatures, in constrast to classical signatures, allow for high dimensional market dimension and provide features on the same scale. We do not contribute to the theory of Randomized Signatures here, but rather present our empirical findings on portfolio selection in real world settings including real market data and transaction costs.
\end{abstract}


\author[add1,add2]{Erdinç Akyildirim}
\ead{erdinc.akyildirim@bf.uzh.ch}
\author[add1]{Matteo Gambara}
\ead{matteo.gambara@gmail.com}
\author[add1]{Josef Teichmann}
\ead{josef.teichmann@math.ethz.ch}
\author[add1]{Syang Zhou}
\ead{syang.zhou@math.ethz.ch}

\address[add1]{Department of Mathematics, ETH, Zurich, Switzerland}

\address[add2]{Department of Banking and Finance, University of Zurich, Zurich, Switzerland}




\maketitle

\section{Introduction}

 Optimal portfolio construction is one of the most fundamental problems in quantitative finance. It refers to selecting and allocating assets to achieve a balance between risk and return. An optimal portfolio aligns with an investor's specific objectives, risk tolerance, and time horizon. In that sense, optimal implies achieving the best trade-off between expected return and risk based on the fact that different investors will have different optimal portfolios depending on their unique goals and risk tolerances. Notice that a priori neither the precise optimization problem nor the underlying model for the evolution of the market are known to the investor. The former needs a quantification of risk tolerance, time horizon, the latter needs an estimation of model parameters.
 
 There are several fundamental methods for constructing an optimal portfolio. Modern Portfolio Theory (MPT) which is developed by Harry Markowitz in his seminal work \citep{markovitz1959portfolio} provides a foundational approach given a model. MPT uses mean-variance optimization to construct the portfolio that maximizes the expected return while minimizing the associated risk (expressed in terms of variance or standard deviation), or, equivalently, minimizes the risk for a given level of expected return.

 The Capital Asset Pricing Model (CAPM) \citep{treynor1961market, treynor1961toward, sharpe1964capital, lintner1975valuation, mossin1966equilibrium} is another cornerstone in portfolio management which helps investors determine the expected return of an investment, particularly for individual stocks or portfolios of stocks. The Capital Market Line (CML) and Security Market Line (SML) represent portfolios derived from MPT. The former shows the relationship between risk and return for a portfolio of all possible investments and the latter illustrates the relationship between the expected return and the systematic risk of an individual asset or portfolio of assets. Another method called the Maximum Sharpe Ratio Portfolio finds the optimal weights by maximizing the portfolio's Sharpe ratio, where the Sharpe ratio measures excess portfolio return over the risk-free rate relative to its standard deviation.

 As an enhancement to the traditional MPT, the Black-Litterman model \citep{black1992global} was developed to incorporate investors' subjective views and market equilibrium returns into optimal asset allocation in a portfolio. Similarly, factor-based portfolio optimization is used to improve CAPM by systematically considering factors that are believed to affect asset returns. Recently, \cite{auh2023factor} show that a parsimonious factor model mitigates idiosyncratic noise in historical data for portfolio optimization. They also prove that a combination of the factor model and forward-looking returns improves out-of-sample performance. As opposed to the Black-Litterman model, Risk-Parity distributes portfolio risk equally across assets and does not look at investor views or expected return projections. Especially after the Global Financial Crisis in 2008, Risk-Parity became a widely followed strategy 
 \citep{roncalli2016risk, costa2019risk, bai2016least}.

\par Another recent yet classical approach for optimal portfolio allocation is using Monte Carlo simulations. This involves generating a large number of random scenarios to model the range of possible future returns for different assets. Of course this again is based on a given model. By repeatedly simulating portfolio performance under various market conditions, investors can assess the distribution of potential outcomes and make informed decisions (\cite{detemple2003monte}, \cite{cesari2003benchmarking}, \cite{cong2016multi}.) The following methodology to take non-normality and fat-tailed distributions into account is bootstrapping which involves drawing random samples (with replacement) from historical returns data to estimate the distribution of returns. \cite{fernandes2012combining} propose a portfolio optimization methodology with equilibrium, views, and resampling. Another stream in the literature is Stochastic Portfolio Theory (SPT) developed by \cite{fernholz2002stochastic}. Unlike earlier theories like MPT and the CAPM, which prescribe how portfolios should be constructed given relatively strong model assumptions, SPT is more descriptive in nature assuming less on the underlying model. It aligns closely with actual observed market behavior. In SPT, the normative assumptions that underpin MPT and CAPM are absent, in particular one does not assume knowledge on hard to observe quantities. SPT employs a continuous-time random process, specifically a continuous semi-martingale, to model the prices of individual securities. Additionally, it incorporates processes that account for discontinuities, such as jumps, in its theoretical framework. 

More recently, Machine Learning (ML) approaches have been significantly instrumental in optimal portfolio allocation due to their ability to handle large datasets, complex relationships, and non-linear patterns. One of the pioneering studies, \cite{ban2018machine}, introduce performance-based regularization (PBR) and performance-based cross-validation for the portfolio optimization problem. They show that PBR with performance-based cross-validation is highly effective at improving the finite-sample performance of the data-driven portfolio decision. More specifically, \cite{heaton2017deep}
investigate the application of deep learning hierarchical models in the context of financial prediction and classification tasks. In particular, they show that applying deep learning methods to constructing portfolios can produce more convincing results than standard methods in finance. We refer the reader to \cite{kim2023robustness} and references therein for a review of data-driven methods and machine learning–based models for portfolio optimization. 
Our methodology can be categorized within the broader context of signature methods which hold significant importance within the context of rough path theory. The signature method as an ML technique has found a wide range of application areas. For instance, \cite{lyons2019numerical}  and \cite{lyons2020non} (discrete and continuous time, respectively) demonstrate that signature payoffs can be exploited to price and hedge exotic derivatives non-parametrically in case one has access to price data for other exotic payoffs. The resulting algorithm is claimed to be computationally tractable and accurate for pricing and hedging using market prices of a basket of exotic derivatives. There is a great deal of flexibility as to how signature methods can be applied. \cite{morrill2020generalised} provides a general approach by unifying the variations on the signature method. They establish a standard set of options that serve as a domain-neutral initial reference point. An empirical study on 26 datasets show the competitive performance against current benchmarks for multivariate time series classification. Recently, \cite{bayer2023optimal} 
propose a new approach for solving optimal stopping problems utilizing signature methods. In particular, their method can be used for American-type option pricing in fractional models on financial or electricity markets.

\medskip

In the current paper, we employ the method of randomized signatures which has been introduced to the literature by \cite{cuchiero2021discrete} and \cite{cuchiero2021expressive}. One application of this methodology can be found in \cite{RandomizedSig_Anomaly}, where the authors employ this new technique to deliver state-of-the-art results in anomaly detection for pump-and-dump schemes with cryptocurrencies. 
Thanks to this novel methodology, we make several contributions to the literature. Our first contribution lies in the non-linear drift estimation of assets using randomized signatures.
The goal of this traditional \textit{predict-then-optimize} approach is to forecast the future (expected) returns and, thus, maximize the Sharpe Ratio, as other conventional methods in supervised learning would do (see, for example, \cite{ML_portfolios}).
The novelty and beauty of our framework, though, relies on a non-linear estimator that can (potentially) incorporate the full trajectory of the price processes extracting features, such as volatility, which has been advocated for being mainly path-dependent in \cite{vola_path_dependent}, or autocorrelation or other time-dependencies, that are usually partially disregarded.
Even though drift cannot be precisely inferred, we believe that the use of a non-linear, non-parametric and robust estimator can make a difference in the resulting performance.
\emph{Non-linearity} comes from the fact that (randomized) signatures are able to capture non-linearities and geometric information lying in the path of a stochastic process; \emph{non-parametric} since it does not involve parameters that need to be fit, but rather hyper-parameters (such as the time series length) which are dependent on the framework we want to adapt; and \emph{robust} because it is model-independent.
We show empirically that our trading strategy, based on such estimator, performs well compared to standard benchmark portfolios both under synthetic data and a real-world setting.\\
The use of signature-based strategies on mean-variance portfolios is not yet very widespread in the literature, although two remarkable papers have been published in the last months. 
Compared to the paper by \cite{futter2023signature}, we can also embed exogenous information in the supervised learning framework, e.g.~\emph{market factors}, we consider proportional transaction costs and show how our strategy can be adjusted to still obtain very good performance under this setting.
While in the former paper there is an elegant analytical expression for the optimized mean-variance portfolio involving signatures, this cannot be easily applied in real life for the curse of dimensionality which is inherent to signatures.
In our case, by leveraging the Johnson-Lindenstrauss lemma (see \cite{JL_lemma} and \cite{cuchiero2021discrete}), we can consider several assets (and potentially many market factors) that are normally prohibitive for the standard signature algorithm. 
Another recently published paper is \cite{Cuchiero2023}, where the authors use exact and randomized signatures for stochastic portfolio optimization. 
In their article, the authors show that under certain relatively mild assumptions on the market conditions, the growth-optimal portfolio prescribed by SPT can be regarded as a \emph{path-functional portfolios}, which can be approximated by signature portfolios and, hence, the portfolio weights can be calculated using an optimization on the randomized or truncated-exact signatures. 
In contrast to our approach, the authors do not use the signatures to predict the (expected) returns, but focus their numerical experiments towards an efficient approximation of the growth-optimal portfolio, whereas the focus of the present paper is on the non-linear prediction of mean returns.
In this sense, our article might be of interest for many practitioners as well.


\bigskip

\par The remainder of this paper is organized as follows. Section \ref{sec:methods} gives the theoretical foundation and description of our methodology. Section \ref{sec: transaction costs} introduces transaction costs and how these are modeled, while Section \ref{sec: benchmark} describes the classical benchmark we will employ in the rest of the work. Section \ref{sec:data} describes the simulated and real-world data we are using and, eventually, Section \ref{sec:results} gives a summary of the results.

\section{Methodology}
\label{sec:methods}
In this section, we recap the theory of randomized signature, which describes several efficient regression bases for path space functionals. In Finance, as well as in other areas, it is an important problem to approximate path space functionals efficiently. 
It is one insight of neural network technology that an ungraded regression basis appears to be superior to a regression basis, where the number of basis elements up to grade $N$ depends exponentially on the dimension of the input signal. 
More concretely, classical neural networks perform better than polynomial regression basis, in particular on high dimensional input spaces. 
In this section, we introduce the counterpart of neural networks for signature methodologies. 
We also emphasize the relation to Reservoir Computing, an area of machine learning where random, possibly recurrent networks are used to efficiently construct regression basis on path space. 
In particular, the focus is not on training the weights between the connections of the network, but rather on the training of a static and memory-less readout map, such as a linear regression, between the generated random basis and specific output.
We summarize the notion used in this section in Table~\ref{table:notation}. 

One possible choice for the construction of reservoirs is the use of signatures, such as in \cite{futter2023signature}, which is an infinite dimensional system. Signatures, originating from rough path theory, have rich theoretical backgrounds and proven properties which make them very suitable as features for machine learning purposes. We refer the reader to \cite{chevyrev2016primer} for the discussion of path integrals and their properties together with general information on signatures. However, signature feature suffer from similar disadvantages as polynomials like the curse of dimension or different scales for different signature components. Both can be overcome by Randomized Signatures.

Instead of the previously (fixed) infinite dimensional system, we search for a better alternative to construct a reservoir. We first fix an activation function $\sigma$, a set of hyper-parameters $\theta \in \Theta$, and a dimension $r_d$. 
Then, depending on  $\theta$, we choose random matrices $A_0,...,A_d$ on $\mathbb{R}^{r_d\times r_d}$ as well as shifts $b_0,..., b_d$ such that maximal non-integrability holds on a starting point $x \in  \mathbb{R}^{d+1} $. 
One can tune the hyper-parameters $\theta \in \Theta $ and dimension $r_d$ such that paths of
\begin{equation}\label{eq:random_sig}
    d Z_t = \sum_{i=0}^d \sigma(A_i Z_t + b_i ) \;  dX^i_t, \qquad Z_0 = z
\end{equation}
approximate path space functionals of $(X_s)_{s \leq t}$ via a linear readout up to arbitrary precision. Notice that we typically take $X^0_t = t $. 
The process $(X_t)_{t\geq 0}$ is the driving path along which we compute the randomized signature $(Z_t)_{t\geq 0}$ and can include endogenous and exogenous information as well.

\subsection{Problem statement}\label{sec: problem statement}
In a financial market with $n \in \NN$ stocks ($n \leq d$), prices are denoted here by $ \mathcal{S} = (S^1, \ldots, S^n) $, which are expressed by values of a factor process $X_t= (X_t^1,..,X_t^n)$ for $t \in [0,T]$. 
The meaning of the factor process can be price itself, or log price, or first differences of the log price.

One of the fundamental questions is how to construct an optimal portfolio for a time horizon $T \in \RR$. 
As a possible solution, economic theory suggests choosing preferences and setting up an optimization problem given knowledge about the financial market. 
Already in the context of this simple setting, two immediate downsides appear: neither preferences nor the stochastic model for the financial market are easy to specify.

In our context, we assume that the preferences are known and the stochastic process $\mathcal{S}$ is determined by an It\^{o} semi-martingale with continuous trajectories and we let $[-T_{obs},0]$ to be the observation period and $(s \longmapsto X_s)_{s \in [-T_{obs},0]}$ to be the observed data trajectory.\\

Then the question is to find a predictor for the law of
$dX_t = \mu_t \, dt + \sum_{i= 1}^{n} \sigma^i_t \, dB^i_t$, which are the stocks' factor processes, 
when both $\mu$ and $\sigma^i$ are unknown quantities with full path dependence on the observation $\sigma$ algebra generated by $X$. 
Once all factor processes are considered, the quadratic covariation is calculated as follows:
$$[X,X]_{s,t}(\omega) =  \sum_{i=1}^n \int_s^t  \sigma^i_v \otimes \sigma^i_v dv,$$ where  $\sigma^i_v \otimes \sigma^i_v$ is the instantaneous covariance matrix along $t \longmapsto X_t(\omega)$ which is observable along trajectories of $X$. 
We estimate the instantaneous covariance using the shrinkage covariance estimator described in Section \ref{sec:shrinkage}, which is basically an adjusted sample covariance estimator incorporating prior beliefs.

Next we consider the task of drift estimation (under a mild martingale assumption on the stochastic integral):
$$
E[X_t-X_s \,|\,  \mathcal{F}_s ] = E\left[   \int_s^{t} \mu(X_s) \, ds \,|\, \mathcal{F}_s \right] + 0 
$$
which requires a lot of trajectories in order to estimate 
$$
F_{\Delta}( X_{[0,s]} ) = \mathbb{E} [X_{s+\Delta} - X_s \,|\, \mathcal{F}_s ]  = \int_s^{s+\Delta} \mathbb{E}[\mu_v \,|\,\mathcal{F}_s ] \, dv,
$$
where $F_{\Delta}$ is path space functional, i.e.~almost surely defined on $C([0,s];\RR^{n})$ with values in $\mathbb{R}^n$. The necessity of a large amount of trajectories (only obtainable by long term observations) required in order to capture the drift within a certain interval is obvious and is outlined in the following.\\

This fundamental fact does not depend on the simplicity of the underlying model. 
Thus, for ease of exposition, 
assume that a single stock follows geometric Brownian motion with drift $\mu \in \RR$ and volatility $\sigma > 0$, then we can write 
\begin{equation}\label{eq:gbm1}
    \frac{dS_t}{S_t} = \mu\,dt + \sigma \,dW_t  \; \text{ for } t > 0 \; \text{ and } S_0 = s_0 > 0.
\end{equation}
If we collect observations from this stock at every $\Dt > 0$ in a time duration of $T$ such that $N\cdot \Dt = T$ then we observe $N$ points. In this setting, an (optimal) unbiased estimator for the drift is given by
\begin{equation}
    \widehat{\mu} := \frac{1}{N\, \Dt} \sum_{i=1}^N \frac{\Delta S_i}{S_i} = \frac{1}{T} \sum_{i=1}^N \frac{\Delta S_i}{S_i},
\end{equation}
where $\Delta S_i = S_i - S_{i-1}$.
From Equation \eqref{eq:gbm1}, we can compute the main contribution to the variance of $\widehat{\mu}$ as 
\begin{equation*}
    \Var(\widehat{\mu}) = \frac{1}{T^2} \sum_{i=1}^N \Var(\sigma \, \Delta W_t) = \frac{\sigma^2}{T},
\end{equation*}
which obviously implies a standard deviation of $\nicefrac{\sigma}{\sqrt{T}}$.
Consequently, if we want to compute  95\% confidence interval to get the value of the drift in a 1\% window (i.e.~$\pm 0.5\%$), then we have
\begin{equation*}
    q(\alpha)\, \frac{\sigma}{\sqrt{T}} \leq 0.5\%,
\end{equation*}
which leads to $T \geq (1.96 \,\sigma / 0.005)^2$. That is, for $\sigma= 20\% $ we should wait for more than 6'146 years in order to get an unbiased estimator of the drift with 1 \% significance. This is clearly unfeasible.\\

\subsection{Drift Estimation using reservoir computing}\label{sec: drift estimation}
We start our analysis with a set of $n$ stocks, $\mathcal{S} = (S^1, \ldots, S^n)$, such that  $S^j_t$ represents the price of stock $j$ for day $t$. After normalizing each stock's price by its initial value, we then take the first differences of log prices to obtain log-returns which are denoted by $\LR_t^j = \log(S_t^j) - \log(S_{t-1}^j)$. In our model, the evolution of randomized signatures of these log-returns is determined by
\begin{eqnarray}\label{eq: ode_reservoirs}
d R_t = \sum_{i=0}^d \sigma(A_i R_t + b_i )  
\;  dX^i_t \,, \qquad R_0 \sim \mathcal{N}(0, I_{r_d}),
\end{eqnarray} 
where $\sigma$ is used as activation function and $d+1$ represents the input dimension, with $n \leq d$ (see \cite{RandomizedSig_Anomaly} or \cite{cuchiero2021discrete} for more information). 
Here we have followed the standard approach by setting time as the $0$\textsuperscript{th} 
dimension. 
Moreover, $A_i \in \mathbb{R}^{r_d \times r_d}$ are linear random projection operators where each entry follows a normal distribution with mean $r_m$ and variance $r_v$, and the entries of the random biases $b_i \in \mathbb{R}^{r_d}$ follow a standard normal distribution. 
Note that $A_i$ and $b_i$ are only generated once per ``simulation'', which means that they will be constant for different instants of time, but vary across different experiments. 
Given the stochastic nature of such experiments, we will then take the average to obtain the mean-behavior. 
The initial value of the randomized signature $R_0$ is drawn randomly as a standard normally distributed vector in $\RR^{r_d}$.
Finally, $X$ represents the leading increments in Equation~\eqref{eq: ode_reservoirs} and it contains the log returns of the prices, $\LR_t$, or the log-prices themselves, $\log S_t$. 
Note that the information content is the same independent of choosing $\log S_t$ or $\LR_t$ as input. 
In the following, we follow the usual convention in finance and we use $\LR_t$ for our experiments to make our results comparable to the standard benchmarks, but we experimented with $\log S_t$ and obtained similar results\footnote{Results are available from the authors upon request}. 
Finally, the data is augmented using additional inputs as described in Section \ref{sec: data augmentation} and $X^0 =t$ so that we end up with $X = \left(t, LR^1, \ldots, LR^n, a_1, \ldots, a_{d-n} \right) \in \RR^{d+1}$, where $a^i$ denote the additional inputs described in Section \ref{sec: data augmentation}, as our final input for computing the reservoirs.
The general notation is also grouped in \ref{table:notation}.\\

\textit{Notation:} in the following, we will use type-written characters to denote indices (integer numbers) on a time grid and normal characters to denote the associated instant of times. 
For example, if we uniformly discretize the time horizon with a step of length $h$, we denote it by using $T = \tpT \cdot h$, where $T$ denotes the instance of time and $\tpT$ the time index.\\

For any time $\tpt \in \{\tpt_\tpw, \dots, \tpT\}$\footnote{We start from $\tpt_\tpw$ because we need at least $\tpt_\tpw$ data-points to have a valid data-window to use for our algorithm.} we calculate the randomized signature $R_\tpt$ using the previous $\tpt_\tpw$ log-returns as an input. 
That is done so that the input data $X$ for our prediction method is always the same as for the momentum benchmark \ref{sec: benchmark}.
Hence, in the typical case where $X_t \equiv \LR_t$, any single reservoir entry $R_\tpt$ depends on $\{\LR_{\tpt-\tpt_\tpw}, \dots, \LR_{\tpt-1}\}$ and on hyper parameters $(r_d, r_m, r_v)$. Let
\begin{equation}\label{eq: Reservoir_function}
    \mathcal{F}^{r_d, r_m, r_v}: \mathbb{R}^{\tpt_\tpw \times d} \longrightarrow \mathbb{R}^{r_d}
\end{equation}
be a numerical scheme solving the Equation~\eqref{eq: ode_reservoirs} such as forward Euler scheme.

It is important to note that the evolution of the randomized signature is determined by the increments of time series $X$. 
As we mentioned above, the standard approach is to add a time-dimension to the log-returns and use them in Equation~\eqref{eq: ode_reservoirs}. 
Our numerical experiments also showed that normalizing returns by their first value yields better results\footnote{Note that normalization on the stock prices becomes redundant if the returns are normalized. 
However, in our experiments we still work with the normalized price data because the same time series is then used for the benchmark algorithms.}.\\

In principle, it is possible to add new information to the simple log-returns different from time to enhance the prediction accuracy. 
For instance, we experimented adding  realized standard deviations and obtained similar results as including time as an extra dimension\footnote{Results for the standard deviation are available from the authors upon request}. 
The only shortcoming of this approach is that we increase the dimension of the input time series with other $d$ new time series. 
However, note that this does not create any problem in our methodology since we exploit random projections which are easily generated and do not suffer from any curse of dimensionality.

Our goal is to learn stock behaviour from past observations, hence we apply a supervised  learning algorithm. Since a minimum amount of data is necessary for learning, we split the dataset in two different groups: the ``burn-in'' set is denoted by $I_{\text{burn}}(t_s)$, and the train set is denoted by $I_{\text{train}}(t_s)$ where the  $t_s$ is the time of separation. As Table~\ref{table:notation} shows, the two subsets $I_{\text{burn}}(t_s)$ and $I_{\text{train}}(t_s)$ are always non-overlapping and connected regions of the ordered interval $\{\tpt_{\tpw}, \dots, \tpT\}$. Hence $I_{\text{burn}}(t_s)$ will be a static region always used in its entirety and $I_{\text{train}}(t_s)$ will be a gradually expanding training set.

As it is standard for reservoir systems, we only need to learn the read-out map from the randomized signature controlled by $X$ given by Equation~\eqref{eq: ode_reservoirs}. 
In our numerical application, we try to predict one day ahead in the future. 
The read-out map is obtained from a classical Ridge regression that involves a regularization parameter $\alpha = 10^{-3\,}$\footnote{We chose this particular value for the smoothing parameter, as in our experiments, this results in regression coefficients in a reasonable range.}. 
This map is obtained for every time and it is computed on input spaces of increasing dimension. 
In other words, at each time instant, we will have a new randomized signature element in $\RR^{r_d}$ added to the previous ones. 
The increase in the sample space can be considered \emph{a priori} as a downside of the method, however we emphasize that this does not represent a practical obstacle for us  given the linear nature of the operated regression.

In our numerical example, we chose $t_s$ such that $t_s = \frac{T}{10} > t_w$ which in turn implies that we only look at $t$ in the interval $\{\tpt_\tps+1, \dots, \tpT\}$.
Once $\tpt$ is fixed, then we compute the randomized signature of $X$ over the interval $\{\tpt_\tpw, \dots, \tpt\}$ and map it against the log-returns $\{\LR_{\tpt_\tpw+1}, \dots, \LR_{\tpt+1}\}$. The readout which is an $L^2$-regularized linear regression is deployed on the input sample $R_{\tpt+2}$ to obtain the prescribed output $\LR_{\tpt+2}$. The process is then repeated for all $\tpt$ until $\tpT-1$ is reached.

\subsection{Covariance Estimation}\label{sec:shrinkage}
In the previous section, we discussed our methodology for estimating the expected returns. The remaining ingredient required to apply Markowitz portfolio optimization is an estimation of the covariance matrix of returns. It is well known that the use of a standard covariance matrix estimator given by 
\begin{equation}\label{eq: covariance estimator standard}
        \hat{\Sigma}_\tpt =\sqrt{\dfrac{1}{t_w-1} \sum_{i=\tpt-\tpt_\tpw}^{\tpt-1}\left(R_\tpi - \overline{R_\tpt}\right)^2}
\end{equation}
is not appropriate for this task. The main reason is that the parameter space can be too large compared to the sample size which turns into an unstable and unreliable estimator \cite{michaud1989}.

To overcome this difficulty, various techniques have been proposed like shrinkage estimators \citep{Stein1986}. 
All linear shrinkage estimators follow a Bayesian approach, where the sample covariance matrix, Equation \eqref{eq: covariance estimator standard}, is shifted to a prior belief about the covariance structure, for instance, a diagonal matrix containing only the sample variance of the single stocks. 
The resulting matrix can be seen as a convex combination of the sample covariance and a target matrix, usually a scalar value multiplied by the identity matrix.
In the limit case, when the number of samples grows to infinity, the shrinkage estimator converges to the sample covariance estimator, which is in line with the intuition that we slowly move away from our prior belief the more data we receive.
\cite{Ledoit2017} propose an improvement to the simple linear shrinkage estimator by using a nonlinear shrinkage estimator, 
which allows to shrink with a different intensity for the different eigenvalues of the sample covariance matrix (while keeping the same eigenvectors) and, most importantly, without requiring any prior knowledge on a target covariance matrix.
Furthermore, they prove the optimality of their estimator inside the class of rotation-equivariant estimators (thus, estimators that do not modify sample covariance eigenvectors).

Note that both linear and nonlinear shrinkage techniques give rise to positive definite, hence invertible, covariance matrices. 
The use of shrinkage makes the covariance estimator more robust in general, which means that it will vary less across time.
As covariance estimation is not the main focus of this article, we refrain from formally stating the assumptions and the theoretical results and refer to \cite{Ledoit2017} for a detailed description of the non-linear shrinkage estimator.
In our experiments, we use an implementation of the estimator in \textbf{\textsf{R}}, a statistical programming language (package provided by \cite{nlshrink}).

\subsection{Portfolio weights generation}\label{sec: weights generation}
To assess the quality of our predictions, we construct the maximum Sharpe ratio portfolio first proposed in \cite{sharpe}. 
Additionally, we impose holding constraints on the weights by not allowing short selling and additionally limiting the maximum allowed weight per single asset at 20\% of the total assets. 
Hence the Markowitz optimisation problem that we solve takes the following form for any time step:
\begin{equation}\label{eq: maxsharpe}
\begin{cases}
\underset{w}\argmax \dfrac{w^\top \hat{\mu} - r_f}{\sqrt{w^\top\,\hat{\Sigma}\,w}}, \\
\sum_i |w^i| = 1, \\
0 \leq w^i \leq 0.2.
\end{cases}
\end{equation}
Here $\hat\mu$ is given by the estimator described in Subsection~\ref{sec: drift estimation}, $\hat{\Sigma}$ is the sample covariance matrix obtained after by non-linearly ``shrinking'' the sample unbiased estimator as described in Subsection~\ref{sec:shrinkage}, and $r_f$ is the risk-free rate. 
For the sake of simplicity, we do not consider transaction costs for our experiments using simulated data and compare the annualized returns and the annualized Sharpe ratios of our methodology to the benchmark portfolios described. For real-world data, we compare the annualized returns and the annualized Sharpe ratios under different levels of transaction costs.


\subsection{Additional input information \label{sec: data augmentation}}

As we mentioned before, we can in principle feed our methodology with any other kind of available information coming from the market or user generated data with the available information.
For instance, we try our algorithm with the inclusion of the following:
\begin{itemize}
    \item Random generated portfolio\\
    The idea is to add new artificial time series obtained by fixing some randomly chosen weights at starting time throughout the entire process.
    The intuition is that these new information should be able to increase the signal-to-noise ratio\footnote{For stochastic quantities, the signal-to-noise ratio (SNR) is defined as the ratio between the second moments of the random variables describing the signal $\fS$ and noise $\fN$, that is $\text{SNR} = \EE[\fS^2]/\EE[\fN^2]$.} and help to identify the stocks based on their first moment (drift).
    
    \item Volatility of the mean-returns\\
    In this case, we computed the mean of all returns and then the volatility of such an average.
    The inspiration comes from the fact that in signature applications it is often advisable to add another time series obtained as a transformation from the previous one, namely applying the so-called \emph{lead-lag transformation} which allows to take into account the quadratic variation of the process (see \cite{Gyurko2014ExtractingIF} on this topic).
    
    \item Volatility of each stock's return\\
    The idea is similar to the previous one, but this time entails computing the volatility of each stock separately, avoiding information lost because of the initial average across stocks.
    
    \item Future contract prices on VIX\\
    We increase the time series with the ticker \texttt{UX1} which refers to the first-month VIX futures closing price to consider the possibility of investing in VIX-related products when the volatility of the markets increases since it is not possible to trade on the VIX index directly. In practice, our algorithm invested in VIX futures only 5 days, without bringing a substantial increase in the Sharpe Ratio.
\end{itemize}

Note that adding all this information does not cause any bottleneck for our algorithm because we rely on a random basis and the training only concerns the linear readout.

\section{Transaction costs}\label{sec: transaction costs}
In our experiments with S\&P500 market data, we additionally investigate how the introduction of proportional transaction costs influences the performance by introducing transaction costs via
\begin{equation}\label{eq: transaction cost}
    TC_t = \lambda \sum_i|SH^i_t - SH^i_{t-1}|,
\end{equation} 
where $\lambda$ is the proportional transaction cost and $SH_t^i$ is number of shares in the portfolio in stock $i$ at time $t$. 
Hence, we consider costs when both buying and selling stocks.
The portfolio value is adjusted by subtracting the transaction cost at each time step. 
As our methodology described so far is purely based on drift and covariance estimation to optimize the Sharpe Ratio, portfolio weights can vary widely between trading days leading to bad outcomes in a trading environment that includes transaction costs.

To lower transaction costs in our methodology, we post-process the portfolio weights to make our trading strategy trade less and only on days with a strong trading signal. We achieve this by first smoothing the weights by using a moving average of the past weights and introducing a threshold $\tau$, which has to be breached before trading occurs.
Specifically, given non post-processed portfolio weights $w_t$, predictions $\hat{R}_t$, and stock prices $S_t$, we update the portfolio shares held during the next trading period in the following way:
\begin{equation}\label{eq: adjustment factor}
    \widetilde{SH}_{t+1} = \begin{cases}
    \frac{w_{t+1} P_t}{S_t} & |\hat{R}_{t+1} - \hat{R}_t| < \tau,\\
    SH_t, & \text{otherwise},
    \end{cases}
\end{equation}
where $P_t$ denotes the portfolio value at time $t$.
Given this preliminary update, to further reduce trading costs, we take a moving average over the last $k$ days and re-normalize to get the final share value of the next trading period.
\begin{align*}
    \widetilde{SH}_{t+1}^{k} &= \frac{1}{k} \left(\widetilde{SH}_{t+1}\sum_{i=t-k+1}^t SH_t\right) \\
    SH_{t+1} &= \dfrac{\widetilde{SH}_{t+1}^{k} P_t}{\sum \widetilde{SH}_{t+1}^{k} P_t}
\end{align*}
    
For our numerical experiments, following the experiments performed in \cite{Ruf2020}, we range the proportional trading costs between 0\% and 1\%.

\section{Benchmarks' description}\label{sec: benchmark}

In this chapter, we introduce three benchmark portfolios to compare our methodology.

\subsection{Linear Regression Portfolio}
As a direct comparison to our methodology, we compare our results to a portfolio based on a linear estimation of the drift. For this, we use the same parameters as for our methodology and estimate the drift using a linear regression where the past $t_w$ log returns are used at each time step to predict the mean of the next log return. Subsequently, we use the linear predictions in exactly the same way to construct the portfolios.

\subsection{Momentum Portfolio}\label{sec: benchmark_momentum}
Following trends of the market as an investment strategy has existed for a very long time. 
The general hypothesis of momentum-style portfolios is that stock market trends have ``momentum'' i.e. that on average stock prices, which have been going up, continue to go up and vice versa. 
There are many momentum-based strategies, for a classical overview of momentum strategies please refer to \cite{jegadeesh}. 
In more recent work, \cite{bhurst2017} constructed equally weighted combinations of momentum strategies of various time intervals. If during that time horizon, the past excess return is positive, this is considered an up trend and hence a long position is taken. Vice versa, a negative past excess return is considered a downtrend and a short position is taken. 
In their extensive analysis, the authors have shown a robust market over-performance of momentum following strategies even taking into account transaction costs and management fees. 

To make the momentum strategy more comparable to our strategy, we choose a momentum strategy with the same time interval as our strategy. 
Hence, the strategy will be the same as described in Section \ref{sec: weights generation} only with the average return used as estimators instead of the estimators described in Section \ref{sec: drift estimation}.

\subsection{ 1/n Portfolio}\label{sec: benchmark_1overn}
The story behind the 1/n portfolio is worth reading. 
As stated in \cite{Brower2011}, Markowitz was awarded a Nobel prize for having defined the mean-variance portfolio, which is a portfolio that tries to maximize the gain (mean) for a given risk or to minimize the risk (variance) for a given return, as a young economist he decided to invest with a simple rule of thumb, that can be called ``1/n'', that is \emph{allocate your money equally to each of n funds}.
During an interview, he said: 
\say{I thought, ‘You know, if the stock market goes way up and I’m not in it, I’ll feel stupid. And if it goes way down and I’m in it, I’ll feel stupid. So I went 50–50.}
In practice, from a numerical perspective, this translates into having at all times the portfolio weights equal to 1/n, where $n$ is the number of stocks that we pick for our portfolio. 

The popularity of this approach, which does not depend on complex strategies or availability of data, substantially grew when \cite{DeMiguel2009} empirically proved that the 1/n approach could outperform many other approaches on out-of-sample data. The importance of such an approach as a valid benchmark stems from \cite{Ruf2020} as well. 
In this case, the 1/n portfolio is said, for example, to outperform the other portfolios (e.g. the market portfolio and the entropy-weighted portfolio) in case of zero transaction costs. Because of these reasons, we decided to include the 1/n portfolio as a benchmark for our method.

\section{Data }
\label{sec:data}
\par The empirical analysis in this paper is based on both simulated data and real data. Both data sets consist of daily stock prices covering roughly 20 years. The next section gives a brief overview of the data used.

\subsection{Simulated data}
\label{sim_data_section}
For our simulated data, we assume that the dynamics of asset prices are given by
\begin{equation}
\label{sim_prices}
    dS^i_t = S^i_t \mu^i_t \cos\left( 0.3 \cdot \sum_{j=1}^{10}S^j_t \right) dt+  S^i_t \sigma^i dW^i_t,
\end{equation}
where $ i \in \{1, \ldots, 10\}$ and $W_t^i$ are correlated Brownian Motions with correlation matrix $R$. For all $i$, we set $S^i_0 = 100$ and simulate 5040 time steps corresponding to 20 years of daily data. The parameters $\mu$ and $\sigma$ are given by
\begin{align*}
\mu &= \begin{bmatrix}  -0.1 &0.2&-0.25&0.25&-0.35&0.22&-0.45&0.25&-0.6&0.28 \\\end{bmatrix}^\top,\\
\sigma &=\begin{bmatrix}  0.1&0.2&0.25&0.3&0.35&0.2&0.4&0.35&0.5&0.25 \\\end{bmatrix}^\top.
\end{align*}
We aim to simulate stock data with a non-linear drift with this choice of dynamics. Because of the cosine term, the drift of each stock can fluctuate between positive and negative, and hence it is not easy to determine which stocks are the best to invest in. As we can extract the real drift term of the process \eqref{sim_prices}, we can compute the real trend evolution starting at some arbitrary value $S_0$. Hence we can also compare the predicted mean with the real mean of the process as derived from \eqref{sim_prices}.

\begin{figure}[H]
\includegraphics[width=\textwidth]{"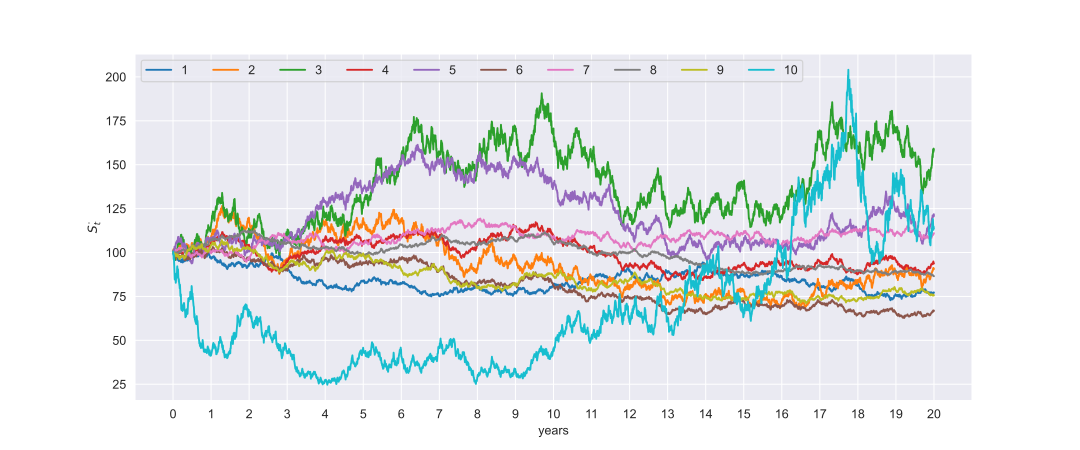"}
  \caption{Paths of the ten simulated stocks obtained as daily observations (for 20 years, every year is made of 252 days).}
  \label{fig: ten_stocks}
\end{figure}

\subsection{S\&P 500 data}\label{sec: snp data description}
For the real data, we collected daily closing prices for a random selection of 50 stocks in the S\&P 500 index from January 2000 until July 2022. 
In the selection of the stocks, we also avoid stocks, which do not have a complete time series for our observation window. Apart from that, we are picking the stocks completely randomly and the list of the used tickers can be found in \ref{sec:list_stocks}. 
Additionally, for our methodology, we perform data augmentation as described in Section \ref{sec: data augmentation}.

Note that this way of selecting stocks introduces survivorship bias, as stocks that have fallen out of the S\&P 500 within our observation period are not picked. The results of our experiments should not be significantly affected, as we are comparing against benchmark portfolios on the same stock selection and which therefore benefit from the same advantages.

We did not perform any experiments targeted at quantifying the impact, as it is not inherently obvious, how to deal with stock prices of varying length within our methodology and we leave this point open for future research.

\section{Results}
\label{sec:results}
In this section we present the empirical results from our algorithm with two different portfolio construction methodologies using both real and simulated prices by the constraint given in Equation \eqref{eq: maxsharpe}.

\begin{table}
\caption{Parameters used for case study}
\label{tab: parameters}
\begin{tabular}{llcc}
\hline
\cline{1-2}
Parameter    & Description & Value \\
\hline
$n_s$      & Number of simulations    & 20      \\
$t_w$    & Days used to calculate reservoir \eqref{eq: ode_reservoirs}          & 22       \\
$t_c$   & Days used to estimate covariance \ref{sec:shrinkage}     & 252      \\
$t_b$ & Fraction of data used for burn-in period & 10\%\\
$r_\alpha$& Regularisation used for ridge regression      & $10^{-3}$       \\
$b_l$ &lower bound for weights & 0\%\\
$b_u$ & upper bound for weights& 20\% \\
\hline
\end{tabular}
\end{table}

 We mainly compare the performance measured in annualized returns and in the Sharpe ratio which we calculate with the following formulas: 
\begin{equation*}
    r_a = (1 + r)^{\frac{252}{t}} - 1,
\end{equation*}
and 
\begin{equation*}
    s_a = \dfrac{r_a - r_f}{\sigma_a},
\end{equation*}

\noindent respectively. As benchmarks for our methodology, we use two separate portfolios. 
First, we compare our methodology with the performance generated using the benchmark log return prediction 
(see Section \ref{sec: benchmark_momentum}) and the same portfolio construction methodology. 
We call this strategy the momentum benchmark.
As a second benchmark, we compare our results with those obtained from using a naive 1/n portfolio meaning that we invest equal weight in each asset re-balance to maintain this equal weighting every tradinig day (Section \ref{sec: benchmark_1overn}). 
In our methodology, there are many specific parameter choices such as the amount of ``burn-in'' data or the regularization constant chosen for the ridge regression. A full table of all parameters used can be found in  Table \ref{tab: parameters}. 
We choose the same parameters for the experiments with the simulated data and the S\&P data.

\subsection{Results with simulated data}

Since we know the dynamics of the stochastic process for the simulated data, we can directly compare our predictions with the true drifts, denoted by $\mu^{*,i}$ for $i \in \{1, 2, \dots, 10\}$, 
\begin{equation}
    \mu^{*,i} = S^i_t \mu^i_t \cos\left( 0.3 \cdot \sum_{j=1}^{10}S^j_t \right).    
\end{equation}
Then we calculate the information coefficient which is the correlation between forecast and realized returns for both our estimates and the true mean. 

\begin{figure}[H]
\includegraphics[width=\textwidth]{"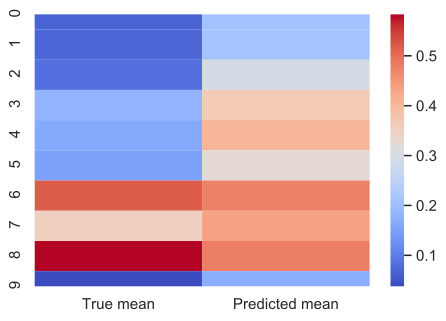"}
  \caption{Information coefficient comparison}
  \label{fig: ic }
\end{figure}

Figure \ref{fig: ic } displays the information coefficient for various stocks. 
As it is clear from the figure, almost all values for our estimates are positive, which provides empirical evidence for the validity of our approach. Furthermore, we can see that for most simulated stock prices, the information coefficient is very low even for the true drift $\mu^{*,i}$, which indicates that for the simulated stocks, one can not perform much better than random picking. 
On average, the information coefficient of our predictions is 3.7\%. 
Although this level of information coefficient may initially appear very low, it's important to consider the high level of difficulty associated with the problem statement, which limits the potential for better predictions.
For instance, \cite{Allen2019} show that an information coefficient of 3.7\% is already sufficient to expect out-performance of the corresponding Markowitz portfolio compared to the 1/n portfolio.

\begin{figure}[H]
\includegraphics[width=\textwidth]{"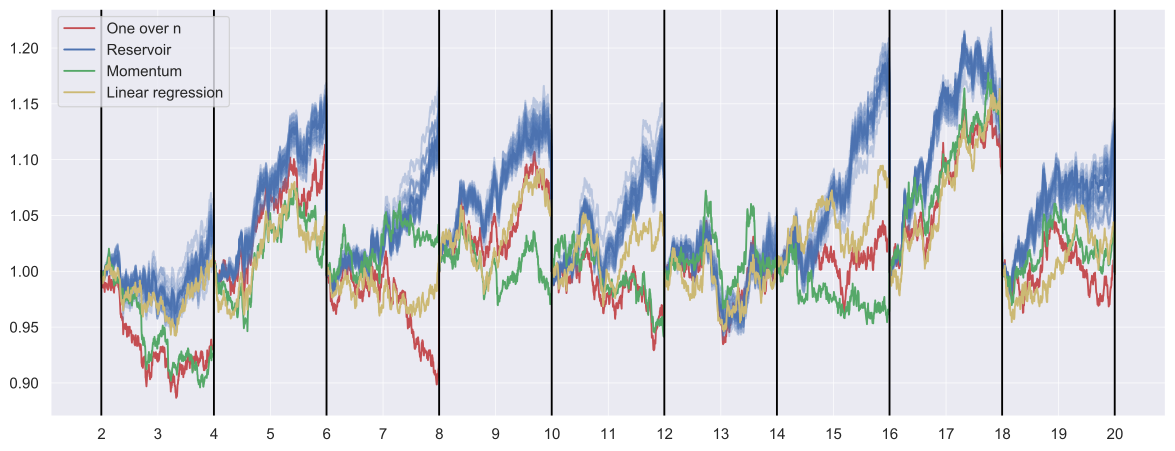"}
  \caption{Performance comparison over simulated data. The portfolio value is normalized to 1 every two years to allow for a better graphical comparison. The first two years were used as a ``burn-in'' period for the algorithm.}
  \label{fig: simulated price comparison }
\end{figure}

Figure \ref{fig: simulated price comparison } displays the performance of our strategy compared to the benchmark portfolios. 
The figure displays the price evolution of portfolios starting at 1 every two years using the methodology described in Section \ref{sec: weights generation}.
The full result, which counts on the compounding effect, can be found in \ref{sec: add_figures} (see Figure \ref{fig: compounded_ratio1}). 
Both pictures start from year 2 since the first two years (10\% of data) have been used as a ``burn-in'' period for the algorithm. 
As we described in the methodology part, we generate $n_s$ predictions for different seeds and use the average of those to generate the price path. 
The bold blue line shows the result of the portfolio constructed in this way. 
Additionally, we also displayed the price paths for each of the $n_s$ predictions (pale blue lines). 
Averaging over the predictions results in a more stable result.

\begin{figure}[H]
\includegraphics[width=\textwidth]{"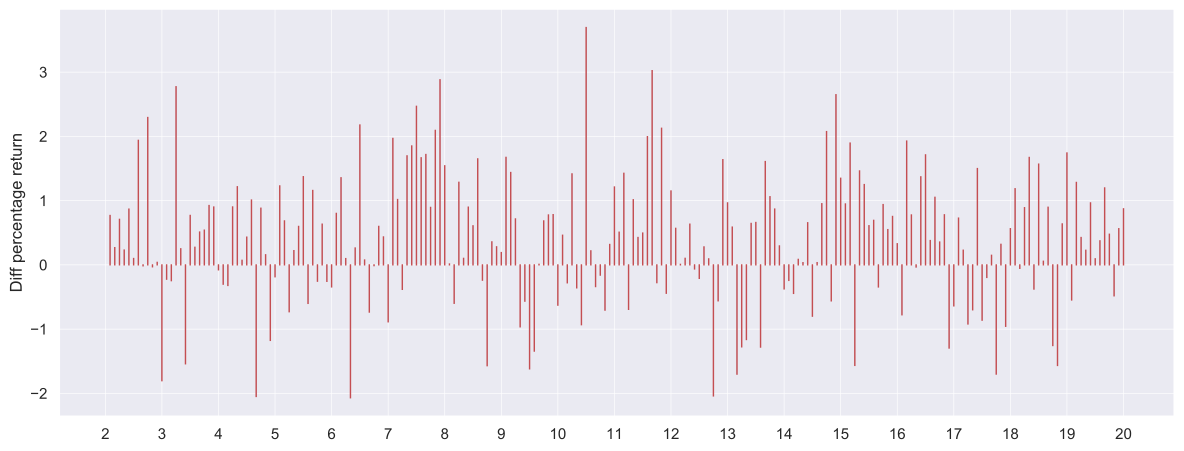"}
  \caption{Performance comparison for simulated data between our strategy and the 1/n strategy. The first two years were used as ``burn-in'' period for the algorithm.}
  \label{fig: simulated_vs1overN}
\end{figure}

To better compare the results, we restart the price paths at 1 every two years, as otherwise, due to the compounding effect of the returns, results are difficult to compare.
This is denoted by a black solid line in Figure \ref{fig: simulated price comparison }.  One can see a clear out-performance of our methodology compared to all benchmarks in most of the path segments. In 7 out of 9 of the two-year segments, our strategy outperforms all benchmarks. In the last temporal segments, our strategy underperforms against two of the benchmark portfolios. 
Due to the low signal-to-noise ratio, it is expected that over-performance cannot always be achieved. Still, the overall results show an over-performance of our strategy across the full simulation period. In particular, if we consider the 1/n strategy as a possible competitor, from Figure \ref{fig: simulated_vs1overN}  we observe that our methodology over-performed the former 67.28\% of the times on monthly returns.

\subsection{Results S\&P 500 data}
As our methodology is dependent on the concrete choice of the reservoir hyper-parameters $(r_d, r_m, r_v)$, we present our results for a grid of different values. 
In our experiments, we choose $r_m \in \{ 0, 0.05, 0.1\}$, $r_v \in \{0.01, 0.03, 0.05, 0.3, 1.0\}$ and $r_d \in \{50, 60, 70, 100\}$, respectively. 
The resulting grid of results compared to the momentum benchmark portfolio can be found in Table \ref{tab:snp full results vs momentum}. 
The corresponding grid of results compared to the 1/n portfolio and the results compared to the linear regression portfolio can be found in Tables \ref{tab:snp full results vs one over n} and \ref{tab:snp full results vs linreg}, respectively. 
The tables show the differences in $r_a$ and $s_a$ correspondingly. As can be seen in the results, our strategy outperforms all benchmark portfolios. As a robustness check for our random signature algorithm, we again observe that positive over-performance numbers can be obtained for all different parameter values being considered.

In addition to supporting our empirical results from the simulated prices, the case study on real-life prices also helps us to recognize the pattern for the optimal parameter values for the highest over-performance statistics. 
First, we observe that for a given level of mean, most of the time it is the case that both over-performance in $r_a$ and $s_a$ values tend to increase with the choice of higher variance values. 
Additionally, we observe that results heavily depend on the concrete choice of the hyper-parameters. 
More specifically, in Table \ref{tab:snp full results vs one over n} one can see that over-performance in $r_a$ ranges from $1.51\%$ for our worst performing model to $6.2\%$ in our best performing model. Naturally, this leads to a significant difference throughout our evaluation period.

\begin{figure}[H]
\includegraphics[width=0.7\textwidth]{"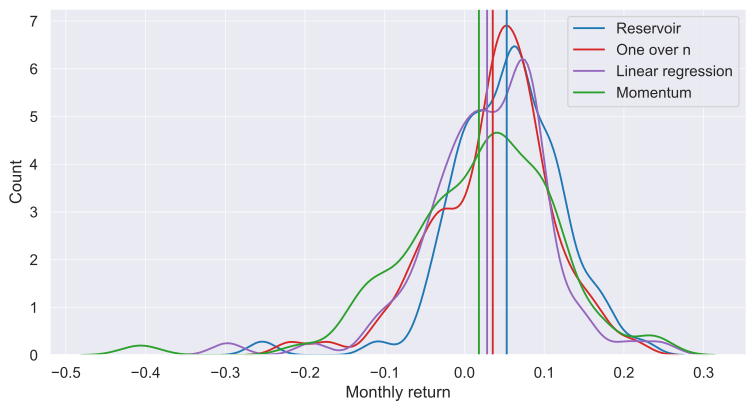"}
  \caption{Comparison over S\&P data of monthly returns provided by the four different strategies. The vertical lines show the means of each of the monthly return distributions.}
  \label{fig: real data}
\end{figure}

We also check the robustness of our results by displaying all $n_s$ results for a given hyper-parameter configuration $(r_d, r_m, r_v)$. 
Figure \ref{fig: real data} shows the monthly returns for a given hyper-parameter configuration $(r_d, r_m, r_v) = (70, 0.0, 0.03)$ for each of the portfolios. As it is clear from the figure the shape of the distribution is similar for each of the portfolios and that of our strategy is indeed slightly higher than the ones for each benchmark portfolio with the 1/n benchmark portfolio being the closest contender.

\begin{figure}[H]
\includegraphics[width=\textwidth]{"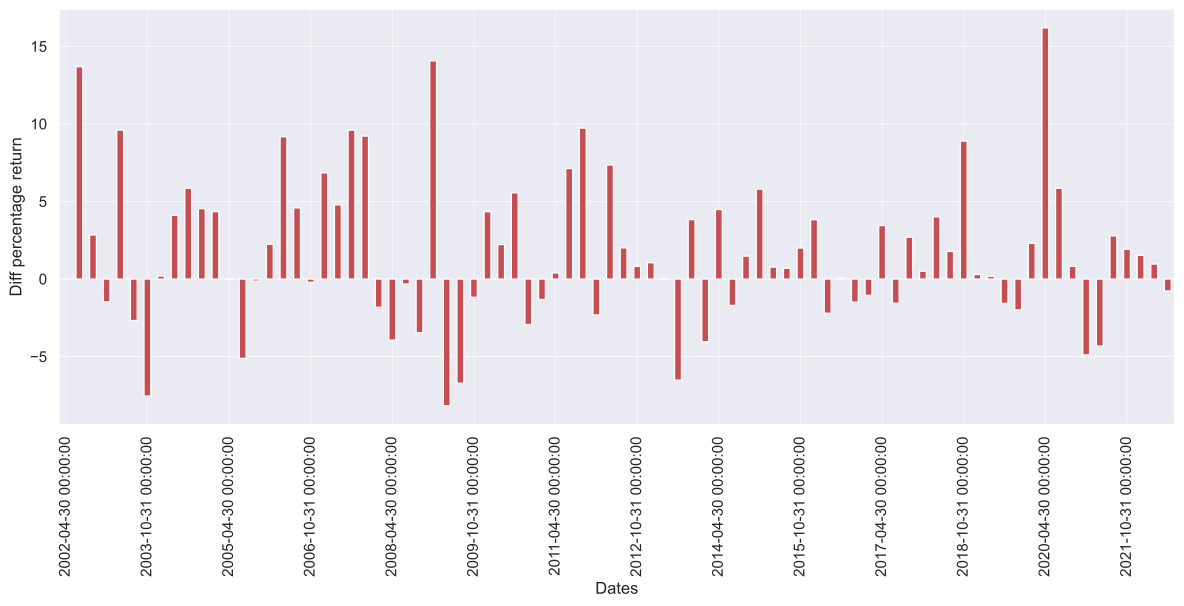"}
  \caption{Difference in quarterly returns compared to 1/n benchmark.}
  \label{fig: 1 over n comparison quarterly}
\end{figure}

For the same hyper-parameter configuration $(r_d, r_m, r_v) = (60, 0.02, 0.03)$, we also check the temporal robustness by comparing the quarterly returns from our methodology with those from the 1/n benchmark portfolio. 
Visual inspection of Figure \ref{fig: 1 over n comparison quarterly} shows that our method's quarterly over-performance does not indicate that our methodology performs better or worse in specific periods. 
Comparing the full period for this specific hyper-parameter configuration, our methodology outperforms the 1/n benchmark in 59.43\% 

\subsection{Comparisons}

Elaborating on our results, we observe a larger over-performance of our portfolio on the real data compared to the simulated data. Specifically, we only get an over-performance of 2.6\% compared to the strongest benchmark portfolio which is less than average over-performance on the real data compared to the strongest benchmark (3.5\%) in Table \ref{tab:snp full results vs one over n}. While this result might seem counter-intuitive at first glance, this behavior can be attributed to the fact that simulated data have lower dimensionality (10 against 50 time series), have a precise correlation structure (which is constant - and invertible! - over time) and, thus, present lower arbitrage possibilities concerning real data.
Note, however, that the covariance matrix used in the portfolio construction was obtained with the same procedures for both simulated and real datasets. We postulate that randomized signatures can better detect arbitrage opportunities when the signal-to-noise ratio is low enough.

\begin{figure}[H]
\includegraphics[width=\textwidth]{"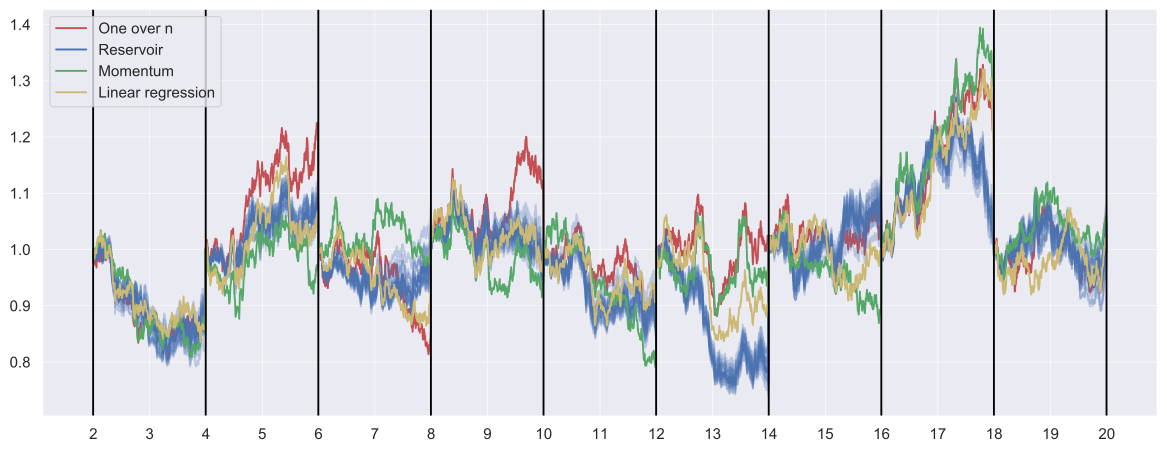"}
  \caption{Performance comparison over simulated data whose Brownian motions' standard deviations were increased by a factor of 2. The portfolio value is normalized to 1 every two years to allow for a better graphical comparison. The first two years were used as a “burn-in” period for the algorithm.}
  \label{fig: performance_ratio2}
\end{figure}

To corroborate our hypothesis, we performed another analysis of the simulated data.
We use the same stochastic simulations for the Brownian motions, but we increased the standard deviation by a factor of 2.
The new simulations can be seen in Figure \ref{fig: ten_stocks_ratio2}.
In this case, randomized signatures do not perform as well as in the previous case.
This can be seen in Figure \ref{fig: performance_ratio2}, where again we re-normalize to the unit value every two years (for the total compounded result, see Figure \ref{fig: compounded_ratio2}).
Under this setting, the 1/n portfolio performs better than our strategy based on randomized signatures.
In particular, our strategy generates monthly returns that are higher than those obtained by 1/n in only 45.16\% of the times (see Figure \ref{fig: simulated_vs1overN_ratio2}).

\begin{figure}[H]
\includegraphics[width=\textwidth]{"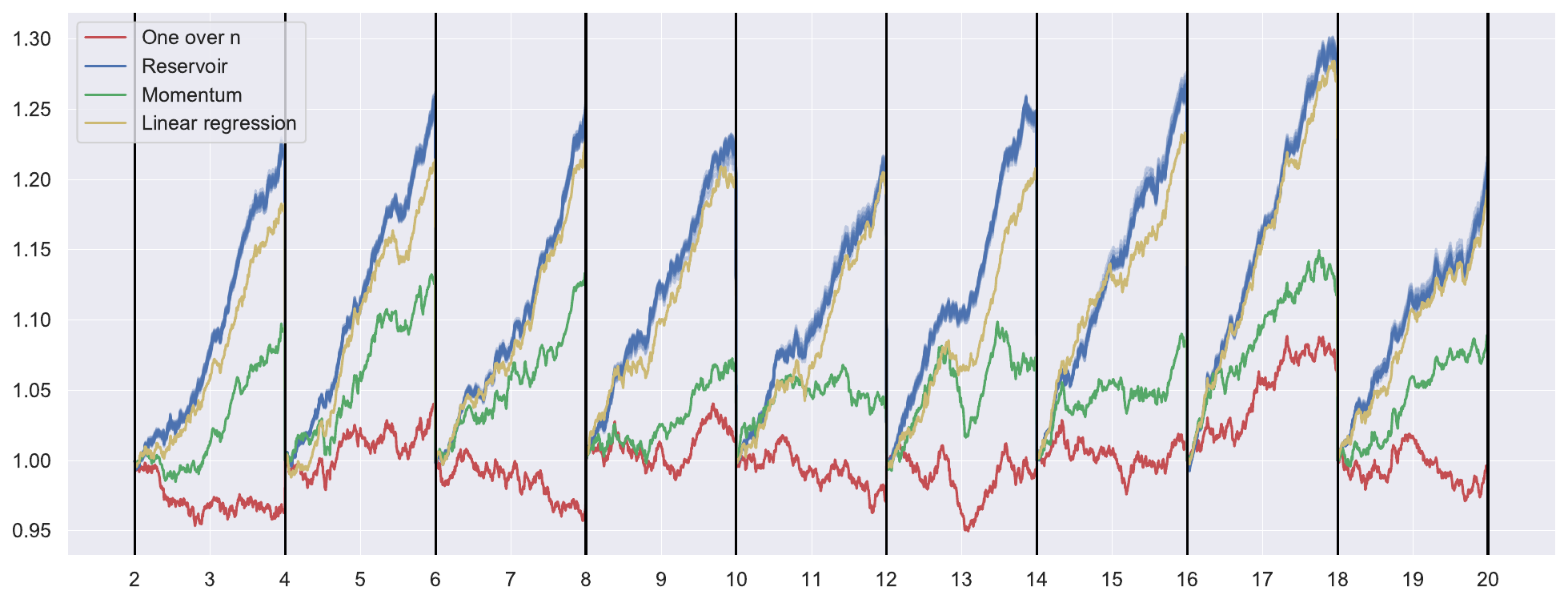"}
  \caption{Performance comparison over simulated data whose Brownian motions' standard deviations were decreased by multiplication with a factor 0.5. The portfolio value is normalized to 1 every two years to allow for a better graphical comparison. The first two years were used as a “burn-in” period for the algorithm.}
  \label{fig: performance_ratio0.5}
\end{figure}

We also try the opposite which is decreasing the standard deviations of the Brownian motions by multiplication with a factor of 0.5 while keeping the same realizations of normal random variables. The purpose is to increase the signal-to-noise ratio and see how the different methodologies perform in this environment.
In this case, the randomized signatures can capture the signal only when the variance $r_v$ used for the construction of the random matrices $A_i$ and random biases $b_i$ for $i=0, \dots, d$ in Equation \eqref{eq: ode_reservoirs} are increased (we used a factor of 4, which corresponds to the reciprocal of the factor used for the standard deviation).
The results are shown in Figure \ref{fig: performance_ratio0.5}. 
As expected, a simple linear regression can also perform well in this low-noise context.
Moreover, we see that our method is consistently better than the linear regression.
This can be also seen in Figure \ref{fig: compounded_ratio0.5}, which displays the entire portfolio history (with a compounding effect).
The monthly return over performance on the 1/n strategy is 94.93\% of the times (see Figure \ref{fig: simulated_vs1overN_ratio0.5}).\\
We also emphasize the fact that this result is only possible by changing the reservoir.
In other words, using the same reservoir as before yields generally lower performance, in line with the 1/n strategy.
It seems that this results from the fact that the random features are not rich enough to filter the correct future returns.

\subsection{Transaction costs}
To make our strategy applicable in practice, we have to consider transaction costs. As mentioned before, we have to adjust our trading strategy in case of an environment with proportional transaction costs. Following the methodology described in Section \ref{sec: transaction costs}, we introduce two additional hyper-parameters $\tau, k$, corresponding to the threshold and the number of days over which we take the moving average respectively.
\begin{figure}[H]
\includegraphics[width=\textwidth]{"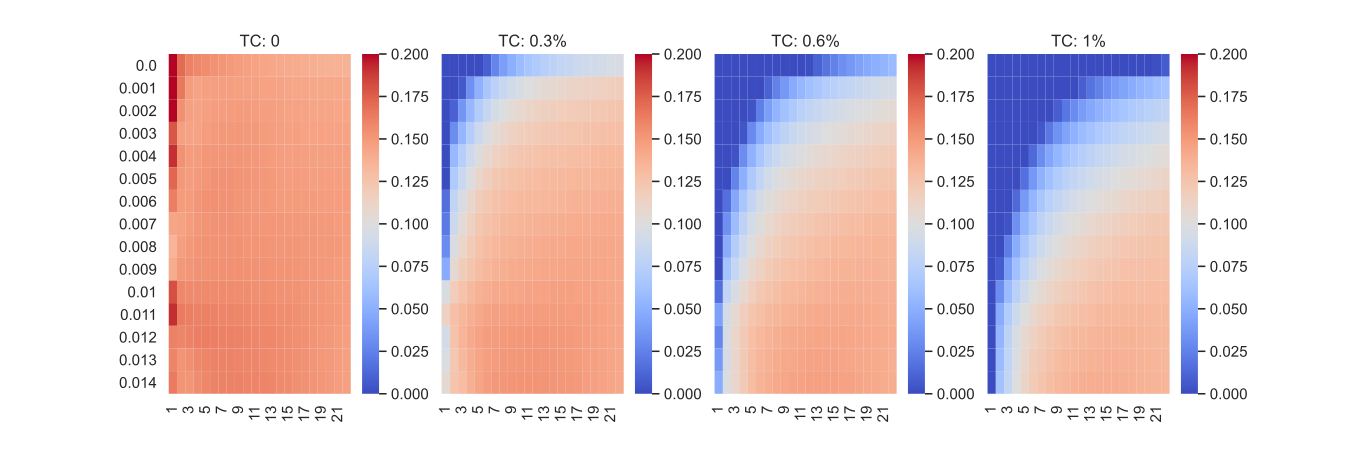"}
  \caption{Performance under four different transaction costs, namely $\{0\%, 0.3\%, 0.6\%, 1\%\}$. On the $x$-axis, there are different values of $k$, window length for the moving average, while on the $y$-axis different values of $\tau$, different threshold values.}
  \label{fig: heatmap}
\end{figure}

Figure \ref{fig: heatmap} shows the effect of different levels of transaction cost on the portfolio performance. 
The $x$-axis shows different values of $k$ and the $y$-axis different levels of $\tau$. 
As can be seen, for the environment without transaction costs, the best hyper-parameter configurations are in the upper left corner, whereas with increasing transaction costs, the best-performing configurations move more and more to the lower right corner. This is in line with the intuition, as with increasing transaction costs, we want to trade less frequently.

Finally, we fix $\tau=1\%$ and $k=5$ and compare our portfolio with our benchmark portfolios, where we adjust all benchmark portfolios but the 1/n benchmark portfolio in the same way as described in Section \ref{sec: transaction costs}. 
The reason why we do not similarly adjust the 1/n portfolio is that for this portfolio we do not have predictions and hence the criterion \eqref{eq: adjustment factor} cannot be calculated.

\begin{figure}[H]
\includegraphics[width=\textwidth]{"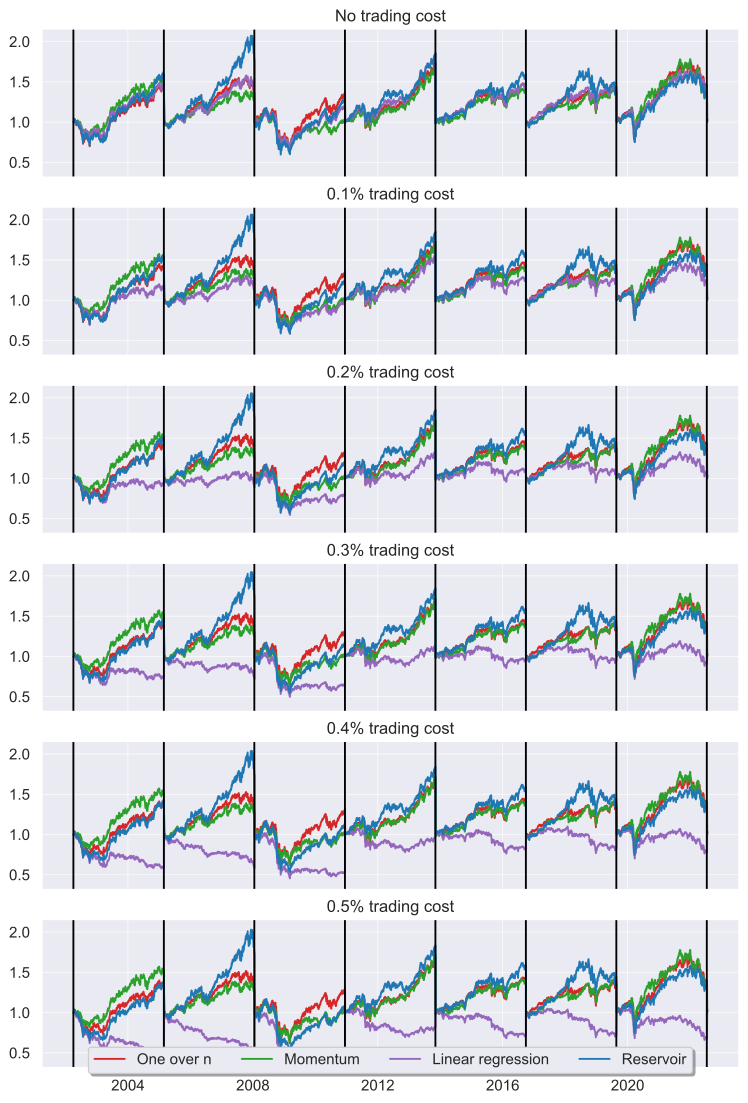"}
  \caption{Comparison under different transaction costs for real data.}
  \label{fig: performance tc}
\end{figure}

\begin{figure}[H]
\includegraphics[width=\textwidth]{"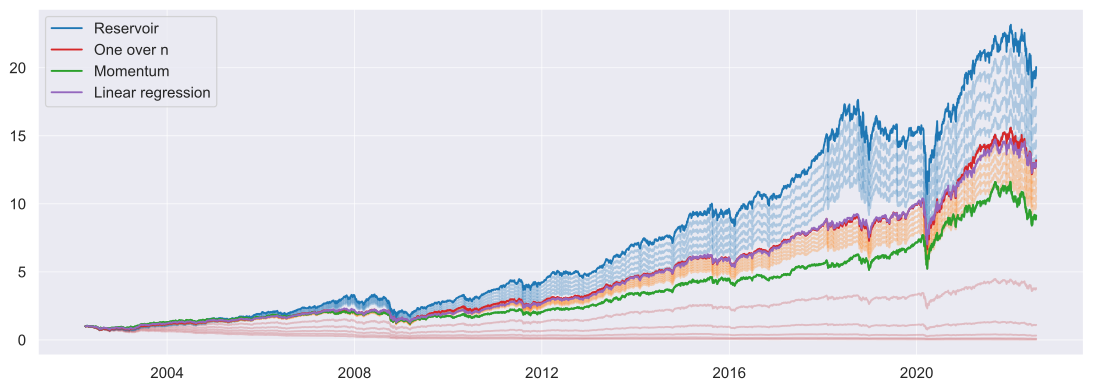"}
  \caption{Comparison under transaction costs, transparent lines showing different levels of trading cost}
  \label{fig: TCPerformance_nocut}
\end{figure}

Figure \ref{fig: performance tc} shows the portfolio paths under varying transaction costs between 0\% and 0.5\%. 
Each of the subplots shows the comparison of the performance of our portfolio vs the benchmark portfolios under different transaction costs. 
Similar to Figure \ref{fig: simulated price comparison }, to be able to better compare the performance, we restart the price paths every three years. The subfigures show that the performance deteriorates in particular for the linear regression portfolio. Also for all other portfolios, the performance deteriorates with increasing transaction costs. Overall, one can see that our strategy outperforms the benchmark portfolio on most of the time segments and trading cost levels. In particular, when we take a look at the performance across the full-time period (see, e.g.,~Figure \ref{fig: TCPerformance_nocut}), one can see that across the full period, our portfolio out-performs all benchmark portfolios for all transactions costs considered.

\section{Conclusion}
\label{sec:conclusion}

\par In this article we investigate whether using non-linear estimators, namely randomized signatures, for return predictions on Markowitz portfolios can outperform a selection of benchmark portfolios. 
We analyze the performance of our portfolio against a portfolio obtained using linear predictions, mean returns, and finally a portfolio with equal weights at each time step. 
We compare the annualized performance and Sharpe ratios on both a simulated data set and a real-world data set. 
Overall, our empirical findings show that random signatures with carefully chosen hyper-parameters can be successfully used as a non-linear and non-parametric drift estimator, in order to optimize portfolio selections for given metrics. Furthermore, our methodology is also robust, as is it does not rely on any model assumptions of the market.

Our experiments on simulated data confirm the intuition that using both linear and non-linear predictions of the mean returns yields better results when improving the signal-to-noise ratio. It also suggests that there is a lower bound for the signal-to-noise ratio under which all predictions break down and the one-over-n portfolio becomes the best-performing portfolio. In our experiments, we show that for signal-to-noise ratios above this lower bound, the portfolio using our methods for non-linear prediction yields the best results. For the real-world data set, our results show a significant over-performance of our portfolio compared to each benchmark portfolio, with the one-over-n portfolio being the strongest benchmark portfolio (see \cite{Ruf2020}). 
Even though the results vary significantly over the hyper-parameter space used for our experiments, we show that our portfolio over-performs for both annualized return and annualized Sharpe ratio even in the worst hyper-parameter configuration.

Last but not least, we consider the impact of transaction costs on our strategy. 
As the original optimization does not take transaction cost into account, it is not surprising that all portfolios show very poor performance, when a sufficiently high proportional transaction cost is used. 
However, we have shown that using an adjustment to our trading strategy using a higher signal threshold can recover the over-performance described in the previous paragraph.

The relationship between the hyper-parameters used for the randomized signatures on the quality of the predictions does not follow a clear pattern and to the best of our knowledge is not well understood. 
For future research, it would be interesting to investigate if it is possible to establish a theoretical link between factors from the input space to successful hyper-parameters configurations. 
Also, for all of our experiments, we have used a single hyper-parameter configuration throughout the whole timeline. 
It would be interesting to see in future research, if our results can be improved upon by varying the hyper-parameter configuration along the timeline potentially using some back-testing strategy.

\bibliographystyle{elsarticle-harv}\biboptions{authoryear}
\bibliography{main}

\newpage

\section*{Appendices}
\appendix

\section{Notation}

\begin{table}[h]
\caption{Notation for the case study}
\label{table:notation}
\begin{tabularx}{\textwidth}{@{}XX@{}}
\toprule
  \bfseries{Dimension parameters} \\
  $n$ & Number of stocks \\
  $d+1$ & Time series input dimension \\
  \bfseries{Indices} \\
  $N$ & Number of observation days \\
  $t_w$ & Rolling window size \\
  $t_s$ & Burn \\
  \bfseries{Index sets} \\
  $\mathcal{N}$ & Set of stocks used for the portfolios\\
  $I_{\text{burn}}(t_s)$ & Initial set $I_{\text{burn}}(t_s) =  \{t_w, \ldots, t_s\}$ \\
  $I_{\text{train}}(t_s)$ & Training set  $I_{\text{train}}(t_s) =  \{t_s + 1, \ldots, T\}$ \\
  \bfseries{Hyper-parameters} \\
  $r_d$   & Reservoir dimension \\
  $r_m, r_v$  & Mean and variance of random projections $A_i$\\ 
\bottomrule
\end{tabularx}
\end{table}


\section{List of stocks from S\&P500 (tickers)}\label{sec:list_stocks}

\begin{multicols}{4}
    \begin{enumerate}
        \item PWR UN
        \item ITW UN
        \item SEE UN
        \item IEX UN
        \item ESS UN
        \item BIIB UW
        \item LUV UN
        \item DD UN
        \item PENN UW
        \item ABMD UW
        \item LUMN UN
        \item BSX UN
        \item DLTR UW
        \item MTD UN
        \item ZBRA UW
        \item CB UN
        \item XRAY UW
        \item TJX UN
        \item AAPL UW
        \item BBY UN
        \item PSA UN
        \item CL UN
        \item REGN UW
        \item NEE UN
        \item DRI UN
        \item PNC UN
        \item BEN UN
        \item MMC UN
        \item DHR UN
        \item TECH UW
        \item DIS UN
        \item ROK UN
        \item L UN
        \item CHRW UW
        \item IPG UN
        \item TSN UN
        \item EFX UN
        \item PCAR UW
        \item EA UW
        \item UNP UN
        \item BKNG UW
        \item TFX UN
        \item WHR UN
        \item NLOK UW
        \item CMA UN
        \item K UN
        \item WST UN
        \item AON UN
        \item VRTX UW
        \item CVS UN
    \end{enumerate}
\end{multicols}

\section{New results}

In the following tables, we show the percentage over-performance (OP) of our model compared to the linear regression benchmark (Table \ref{tab:snp full results vs linreg}), the momentum benchmark (Table \ref{tab:snp full results vs momentum}) and the 1/n benchmark (Table \ref{tab:snp full results vs one over n}).
On the left hand side, it is possible to see over-performances with respect to the annualized returns, while on the right hand side with respect to the Sharpe ratio.

\begin{table}[H]
\begin{tabular}{llrrrr|rrrr}
\toprule
     & {} & \multicolumn{4}{l}{OP annualized return} & \multicolumn{4}{l}{OP annualized SR} \\
     & ReservoirSize &                  50  &    60  &    70  &    100 &              50  &    60  &    70  &    100 \\
Mean & Variance &                      &        &        &        &                  &        &        &        \\
\midrule
0.00 & 0.01 &                 8.74 &  10.04 &   9.27 &   9.73 &            46.99 &  53.27 &  49.93 &  49.73 \\
     & 0.03 &                 8.26 &   9.51 &   8.93 &   9.56 &            44.57 &  50.69 &  48.26 &  49.09 \\
     & 0.05 &                 8.20 &   9.42 &   8.89 &   9.54 &            44.47 &  50.63 &  48.21 &  49.05 \\
     & 0.30 &                10.07 &   8.87 &   8.90 &  11.05 &            53.67 &  47.97 &  47.87 &  56.66 \\
     & 1.00 &                10.15 &   7.87 &   9.38 &  12.20 &            53.43 &  42.83 &  50.02 &  62.51 \\
     \hline
0.05 & 0.01 &                 8.17 &  11.58 &  10.76 &   9.81 &            44.77 &  61.54 &  57.46 &  51.13 \\
     & 0.03 &                 8.69 &  10.73 &   9.79 &   8.26 &            46.82 &  57.26 &  53.26 &  43.45 \\
     & 0.05 &                 8.49 &  11.71 &  10.42 &   9.57 &            45.50 &  61.88 &  56.24 &  50.15 \\
     & 0.30 &                 9.71 &   9.09 &   8.72 &  10.82 &            52.48 &  48.97 &  46.93 &  55.04 \\
     & 1.00 &                 9.66 &   7.83 &   8.53 &  12.04 &            51.28 &  42.62 &  46.07 &  61.90 \\
     \hline
0.10 & 0.01 &                 9.19 &   9.44 &   9.44 &  10.47 &            50.46 &  51.30 &  50.63 &  53.82 \\
     & 0.03 &                 9.73 &  10.32 &  10.84 &  11.52 &            53.31 &  55.23 &  58.53 &  59.39 \\
     & 0.05 &                 9.45 &  10.14 &  10.88 &  11.54 &            51.10 &  54.52 &  58.61 &  60.34 \\
     & 0.30 &                 9.70 &  10.45 &  10.75 &  12.52 &            51.50 &  56.03 &  57.26 &  64.57 \\
     & 1.00 &                 9.25 &   7.83 &   9.24 &  11.94 &            49.63 &  42.48 &  49.40 &  61.27 \\
\bottomrule
\end{tabular}
\caption{\label{tab:snp full results vs linreg} Difference (in \%) of our strategy vs linear regression benchmark}
\end{table}

\begin{table}[H]
\begin{tabular}{llrrrr|rrrr}
\toprule
     & {} & \multicolumn{4}{l}{OP annualized return} & \multicolumn{4}{l}{OP annualized SR} \\
     & ReservoirSize &                  50  &   60  &   70  &   100 &              50  &    60  &    70  &    100 \\
Mean & Variance &                      &       &       &       &                  &        &        &        \\
\midrule
0.00 & 0.01 &                 3.02 &  4.31 &  3.54 &  4.00 &            16.69 &  22.98 &  19.63 &  19.43 \\
     & 0.03 &                 2.53 &  3.78 &  3.20 &  3.84 &            14.28 &  20.40 &  17.96 &  18.79 \\
     & 0.05 &                 2.48 &  3.69 &  3.17 &  3.81 &            14.17 &  20.33 &  17.91 &  18.76 \\
     & 0.30 &                 4.34 &  3.14 &  3.17 &  5.32 &            23.37 &  17.67 &  17.58 &  26.36 \\
     & 1.00 &                 4.42 &  2.14 &  3.65 &  6.48 &            23.14 &  12.53 &  19.72 &  32.21 \\
      \hline
0.05 & 0.01 &                 2.44 &  5.85 &  5.04 &  4.09 &            14.47 &  31.24 &  27.16 &  20.83 \\
     & 0.03 &                 2.96 &  5.01 &  4.06 &  2.53 &            16.53 &  26.96 &  22.96 &  13.16 \\
     & 0.05 &                 2.76 &  5.98 &  4.69 &  3.84 &            15.20 &  31.58 &  25.94 &  19.86 \\
     & 0.30 &                 3.98 &  3.37 &  2.99 &  5.10 &            22.18 &  18.67 &  16.64 &  24.75 \\
     & 1.00 &                 3.93 &  2.10 &  2.81 &  6.32 &            20.99 &  12.32 &  15.77 &  31.60 \\
      \hline
0.10 & 0.01 &                 3.47 &  3.72 &  3.71 &  4.75 &            20.16 &  21.00 &  20.33 &  23.52 \\
     & 0.03 &                 4.01 &  4.60 &  5.12 &  5.80 &            23.01 &  24.93 &  28.24 &  29.10 \\
     & 0.05 &                 3.72 &  4.41 &  5.16 &  5.81 &            20.80 &  24.23 &  28.31 &  30.04 \\
     & 0.30 &                 3.98 &  4.73 &  5.02 &  6.79 &            21.20 &  25.73 &  26.97 &  34.28 \\
     & 1.00 &                 3.52 &  2.10 &  3.51 &  6.22 &            19.33 &  12.18 &  19.10 &  30.97 \\
\bottomrule
\end{tabular}
\caption{\label{tab:snp full results vs momentum} Difference (in \%) of our strategy vs momentum benchmark}
\end{table}

\begin{table}[H]
\begin{tabular}{llrrrr|rrrr}
\toprule
     & {} & \multicolumn{4}{l}{OP annualized return} & \multicolumn{4}{l}{OP annualized SR} \\
     & ReservoirSize &                  50  &   60  &   70  &   100 &              50  &    60  &    70  &    100 \\
Mean & Variance &                      &       &       &       &                  &        &        &        \\
\midrule
0.00 & 0.01 &                 2.43 &  3.72 &  2.95 &  3.41 &            14.60 &  20.88 &  17.53 &  17.34 \\
     & 0.03 &                 1.94 &  3.19 &  2.61 &  3.25 &            12.18 &  18.30 &  15.86 &  16.70 \\
     & 0.05 &                 1.89 &  3.10 &  2.58 &  3.22 &            12.08 &  18.24 &  15.82 &  16.66 \\
     & 0.30 &                 3.75 &  2.55 &  2.58 &  4.73 &            21.28 &  15.57 &  15.48 &  24.26 \\
     & 1.00 &                 3.83 &  1.55 &  3.06 &  5.89 &            21.04 &  10.44 &  17.63 &  30.12 \\
     \hline
0.05 & 0.01 &                 1.85 &  5.26 &  4.45 &  3.50 &            12.37 &  29.14 &  25.07 &  18.73 \\
     & 0.03 &                 2.37 &  4.42 &  3.47 &  1.94 &            14.43 &  24.86 &  20.86 &  11.06 \\
     & 0.05 &                 2.17 &  5.39 &  4.10 &  3.25 &            13.11 &  29.49 &  23.84 &  17.76 \\
     & 0.30 &                 3.39 &  2.78 &  2.40 &  4.51 &            20.09 &  16.58 &  14.54 &  22.65 \\
     & 1.00 &                 3.34 &  1.51 &  2.22 &  5.73 &            18.89 &  10.23 &  13.67 &  29.50 \\
     \hline
0.10 & 0.01 &                 2.88 &  3.13 &  3.12 &  4.16 &            18.07 &  18.91 &  18.24 &  21.42 \\
     & 0.03 &                 3.42 &  4.01 &  4.53 &  5.21 &            20.92 &  22.83 &  26.14 &  27.00 \\
     & 0.05 &                 3.14 &  3.82 &  4.57 &  5.23 &            18.71 &  22.13 &  26.22 &  27.94 \\
     & 0.30 &                 3.39 &  4.14 &  4.43 &  6.20 &            19.11 &  23.64 &  24.87 &  32.18 \\
     & 1.00 &                 2.93 &  1.51 &  2.92 &  5.63 &            17.24 &  10.09 &  17.00 &  28.88 \\
\bottomrule
\end{tabular}
\caption{\label{tab:snp full results vs one over n} Difference (in \%) of our strategy vs 1/n benchmark}
\end{table}

\section{Additional figures}\label{sec: add_figures}

\begin{figure}[H]
\includegraphics[width=\textwidth]{"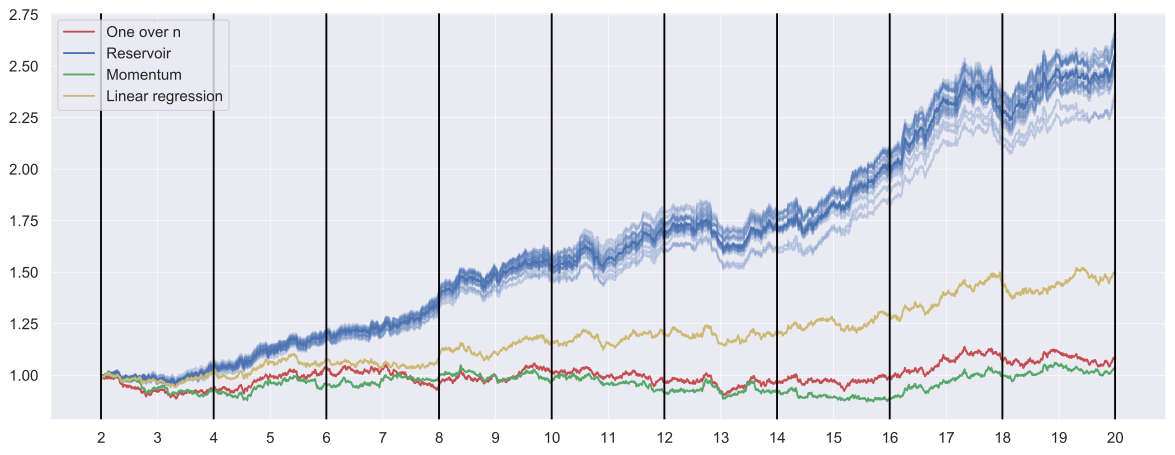"}
  \caption{Performance comparison over simulated data. Compared to Figure \ref{fig: simulated price comparison }, we see the effect of compounding over all years. The first two years were used as ``burn-in'' period for the algorithm.}
  \label{fig: compounded_ratio1}
\end{figure}


\begin{figure}[H]
\includegraphics[width=\textwidth]{"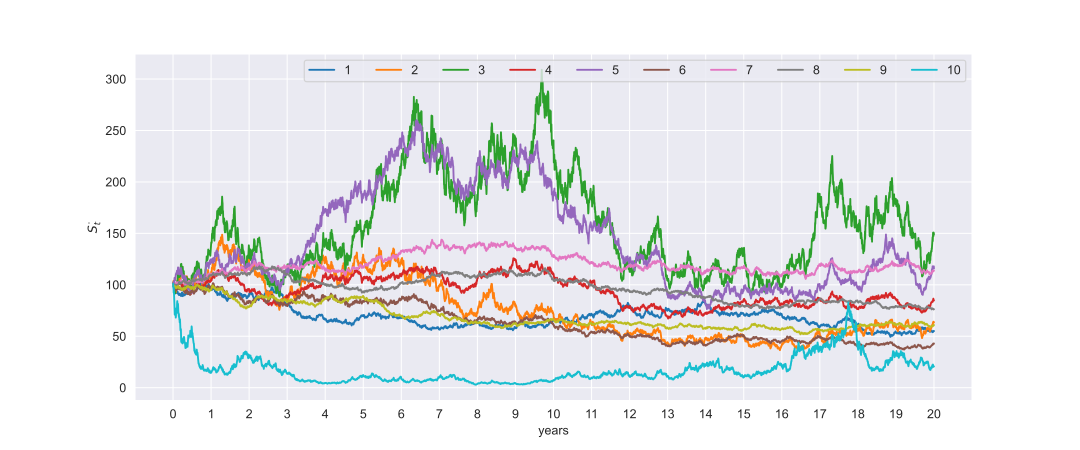"}
  \caption{Paths of the ten simulated stocks obtained as daily observations (for 20 years, every year is made of 252 days).
  Brownian motions' standard deviations have been increased by a factor 2 compared to Figure \ref{fig: ten_stocks}.}
  \label{fig: ten_stocks_ratio2}
\end{figure}

\begin{figure}[H]
\includegraphics[width=\textwidth]{"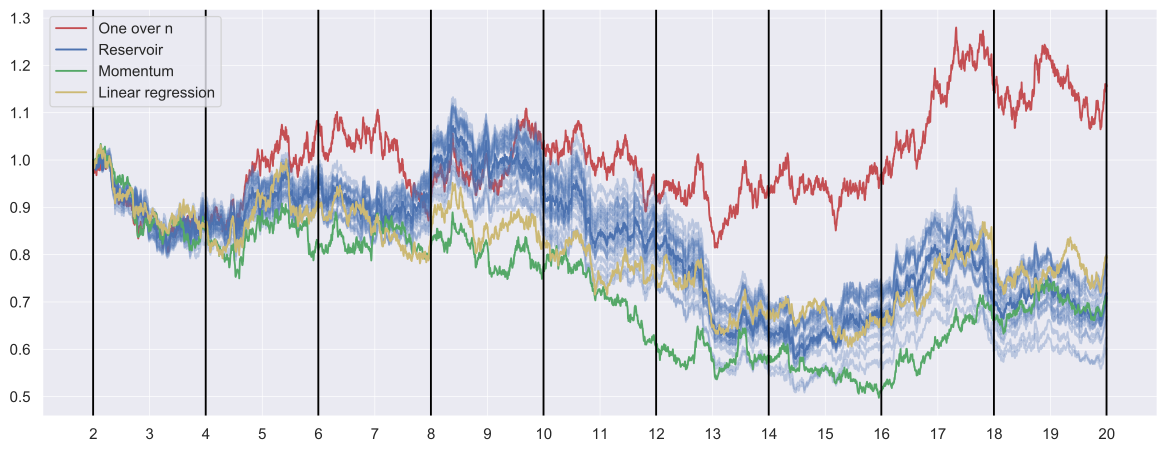"}
  \caption{Performance comparison over simulated data whose Brownian motions' standard deviations were increased by a factor 2. Compared to Figure \ref{fig: performance_ratio2}, we see the effect of compounding over all years. The first two years were used as ``burn-in'' period for the algorithm.}
  \label{fig: compounded_ratio2}
\end{figure}

\begin{figure}[H]
\includegraphics[width=\textwidth]{"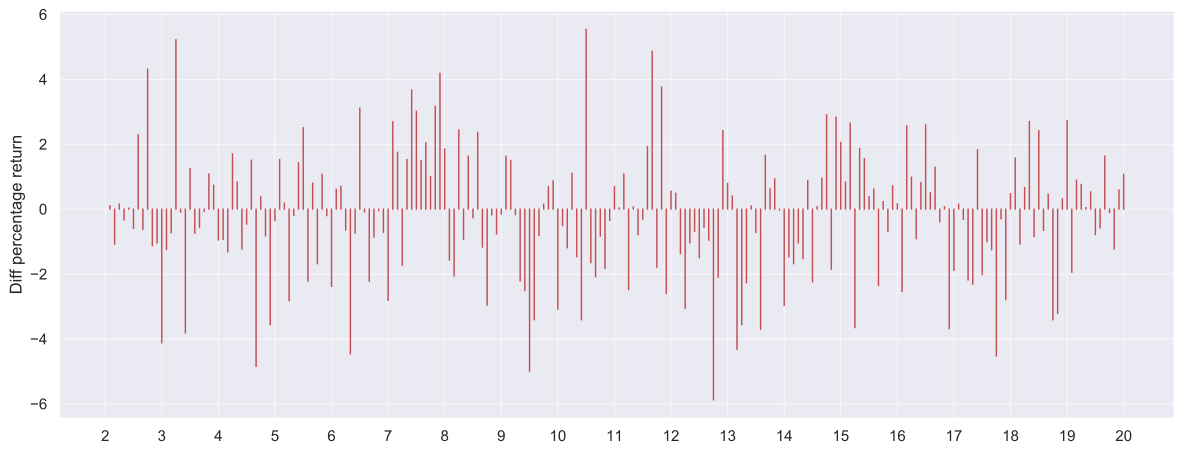"}
  \caption{Performance comparison for simulated data (whose Brownian motions' standard deviations were increased by a factor 2) between our strategy and the 1/n strategy. The first two years were used as “burn-in” period for the algorithm.}
  \label{fig: simulated_vs1overN_ratio2}
\end{figure}


\begin{figure}[H]
\includegraphics[width=\textwidth]{"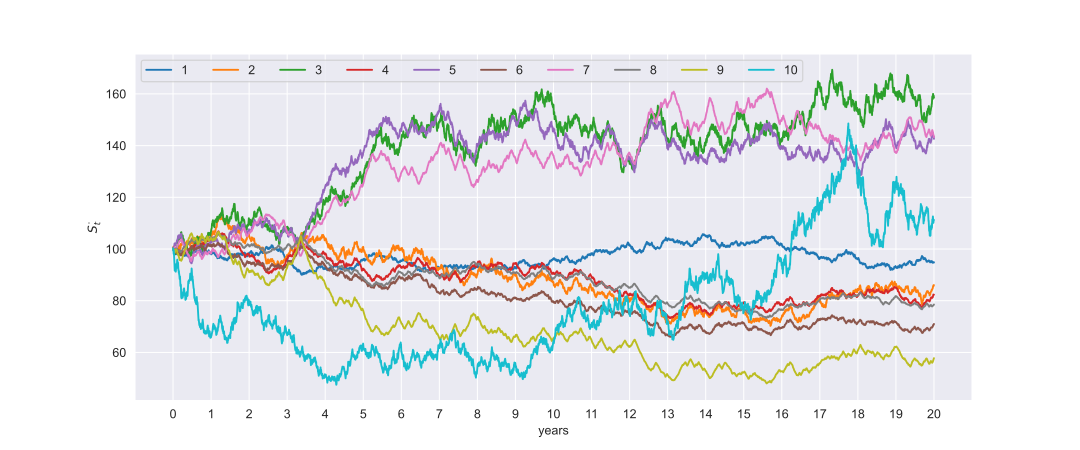"}
  \caption{Paths of the ten simulated stocks obtained as daily observations (for 20 years, every year is made of 252 days).
  Brownian motions' standard deviations have been decreased by multiplication with a factor 0.5 compared to Figure \ref{fig: ten_stocks}.}
  \label{fig: ten_stocks_ratio0.5}
\end{figure}

\begin{figure}[H]
\includegraphics[width=\textwidth]{"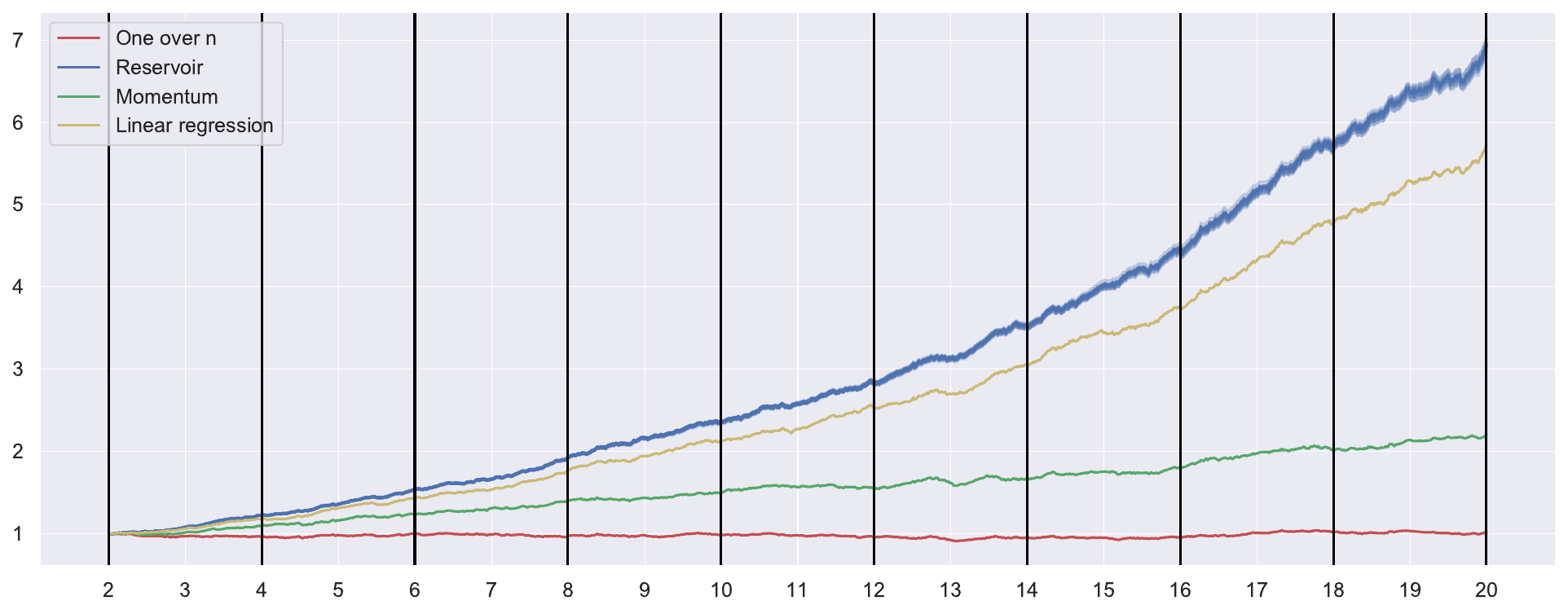"}
  \caption{Performance comparison over simulated data whose Brownian motions' standard deviations were decreased by multiplication with a factor 0.5. Compared to Figure \ref{fig: performance_ratio0.5}, we see the effect of compounding over all years. The first two years were used as ``burn-in'' period for the algorithm.}
  \label{fig: compounded_ratio0.5}
\end{figure}

\begin{figure}[H]
\includegraphics[width=\textwidth]{"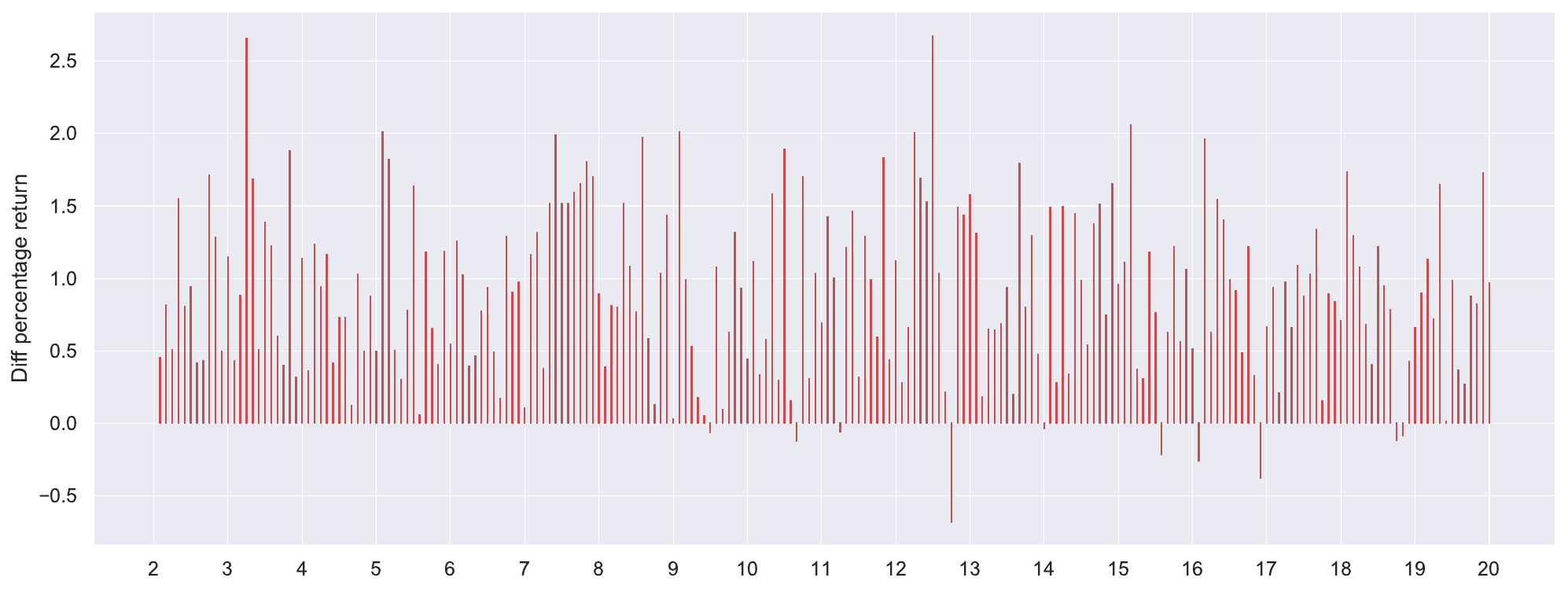"}
  \caption{Performance comparison for simulated data (whose Brownian motions' standard deviations were decreased by multiplication with a factor 0.5) between our strategy and the 1/n strategy. The first two years were used as “burn-in” period for the algorithm.}
  \label{fig: simulated_vs1overN_ratio0.5}
\end{figure}

\end{document}